# Challenges and Experiences of Iranian Developers with MLOps at Enterprise

Mohammad Heydari
School of Industrial and Systems Engineering
Tarbiat Modares University
Tehran, Iran
m_heydari@modares.ac.ir

Zahra Rezvani
School of Computer Science
Institute for Research in Fundamental Sciences (IPM)
Tehran, Iran
zrezvani@ipm.ir

**Abstract—** Data is becoming more complex, and so are the approaches designed to process it. Enterprises have access to more data than ever, but many still struggle to glean the full potential of insights from what they have. This research explores the challenges and experiences of Iranian developers in implementing the MLOps paradigm within enterprise settings. MLOps, or Machine Learning Operations, is a discipline focused on automating the continuous delivery of machine learning models. In this study, we review the most popular MLOps tools used by leading technology enterprises. Additionally, we present the results of a questionnaire answered by over 110 Iranian Machine Learning experts and Software Developers, shedding light on MLOps tools and the primary obstacles faced. The findings reveal that data quality problems, a lack of resources, and difficulties in model deployment are among the primary challenges faced by practitioners. Collaboration between ML, DevOps, Ops, and Science teams is seen as a pivotal challenge in implementing MLOps effectively.

## I. INTRODUCTION

Over the past few years, there has been a significant and rapid growth in acronyms with the "Ops" suffix, which initially started by the merging of development and IT operations. This pairing together of disciplines helped enterprises better define their processes, improve the quality of their output, and operate at much faster speed. With the emergence of artificial intelligence, machine learning, and big data, various enterprises have gained a heightened consciousness of the significance of ModelOps, MLOps, DataOps, and AIOps.

MLOps stands for Machine Learning Operations, is a discipline that focuses on the continuous delivery cycle of machine learning models through automated pipelines. ModelOps on the other hand, is a paradigm used to manage model development from conception to deployment, DataOps provides tools for developing efficient data processing pipelines, while AIOps is an Artificial Intelligence-Driven operations platform that helps automate IT processes such as incident resolution. The intersection of Machine Learning, Model Management and Data Infrastructure in MLOps is an essential element for any organization looking to leverage the power of artificial intelligence. MLOps involves the intersection of machine learning, model management, and data infrastructure to build, test, and deploy machine learning models more efficiently and effectively. By understanding how these three components work together, organizations can better manage their models from conception to deployment. Machine learning is the process of using algorithms and statistical models to automatically improve the performance of a system based on data. It is a key component of MLOps, as it involves building and training machine learning models that can be deployed in production. With MLOps, data engineers can build automated pipelines that facilitate model development and deployment while also allowing for easy monitoring and maintenance.

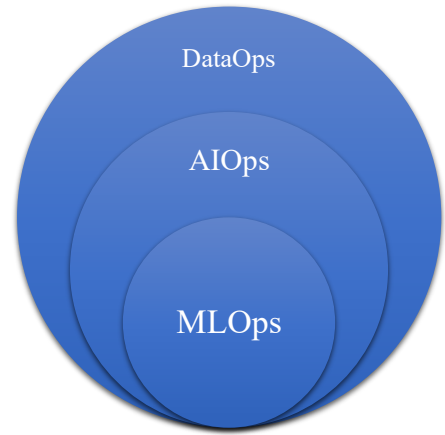

Figure 1 - Intersection of DataOps, AIOps and MLOps [1]

MLOps presents itself as a valuable methodology for the establishment and enhancement of machine learning and artificial intelligence solutions. Through the adoption of an MLOps approach, data scientist and ML engineers can effectively cooperate and expedite the progression of model development and production by implementing the practices of continuous integration and deployment (CI/CD).

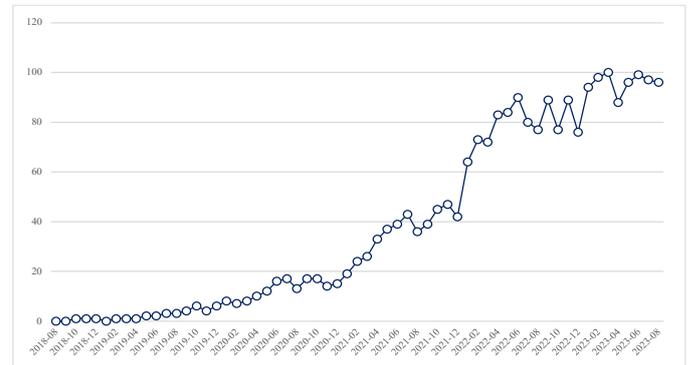

Fig 2 - MLOps Topic Growth on Google Tends from 2018 to 2023 [2]

## II. RELATED WORKS

Both AIOps and MLOps are novel fields, and naturally, they have not yet accumulated a substantial amount of pertinent research and literature. The MLOps inception can be traced back to 2015 by Sculley et al. [3] where they explored several machine leaning specific risk factors to account for in system design, and since then, its development has been notably robust. Sasu Makineth et al. [4] explain the significance of MLOps in data science from a study that engaged 331 industry experts from 63 different countries. Testi et al. [5] review the current scientific researches and propose a taxonomy for clustering scientific reports and studies on MLOps. Symeonidis et al. [6] made an overview of MLOps field by operation and elements definition with focus on challenges, trends and related tools. Renggli et al. [7] explain the importance of data quality for MLOps system while describe how different features of data quality propagate through diverse phase of machine learning development. Cheng et al. [8] review the vision of AIOps, trends challenges and opportunities, particularly with focus on the underlying AI techniques and propose a comprehensive taxonomy of techniques to solve related problems. Lones [9] provides a brief overview of typical errors that arise when working with machine learning and suggests strategies to prevent them. Paleyes et al. [10] reviews published studies of implementing machine learning solutions across diverse use cases, industries, and applications. Their study demonstrates that professionals encounter obstacles throughout the entire deployment procedure. Hewage and Meedeniya [11] studies on examining existing technical challenges associated with software development and deployment within organizations engaged in machine learning projects. Their study describes the availability of MLOps tools for helping software development. Tabassam [12] presents a comprehensive review of MLOps, advantages, challenges progresses and techniques such as frameworks, Docker, Kubernetes and GitHub actions. Ahmed's [13] studies on recognizing trends and valuable understandings to improve the MLOps workflow. His study involves a general MLOps workflow, covering crucial stages such as business problem definition, data ingestion, data preparation, model development, model deployment, monitoring, management, scalability, and governance. Zhengxin et al. [14] made a comprehensive detailed review of the modern MLOps technologies. Calefato et al. [15] investigated on MLOps activities implanted on GithHub with focus on GitHub ACTIONS and Continuous Machine Learning (CML), two workflow's development automation solutions. By identification of related issues, their study results demonstrate integration of MLOps practices in GitHub projects is relatively limited.

## III. MLOps

MLOps represents a new concept and emphasizes how to optimally coordinate data scientists and operations staff for the efficient development, deployment, and monitoring of models since machine learning productization is difficult. In this regard, machine learning, data engineering and software engineering are involved in MLOps paradigm discipline.

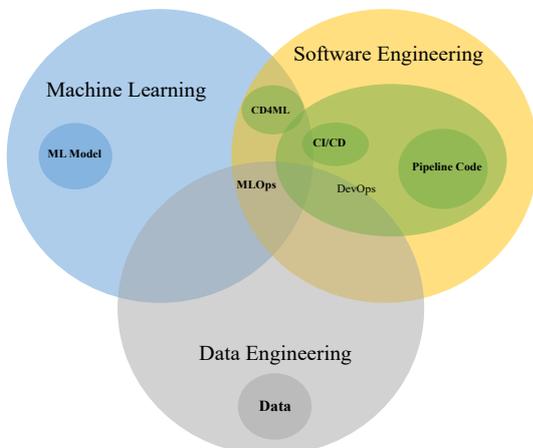

Figure 4 - Disciplines Intersection of MLOps Paradigm[16]

It also necessitates teamwork and transitions between groups, ranging from data engineering to data science and ml engineering. The most important roles in MLOps paradigm are ml engineer, MLOps engineer, DevOps engineer, data engineer, backend engineer and software engineer. MLOps culture includes the following practices:

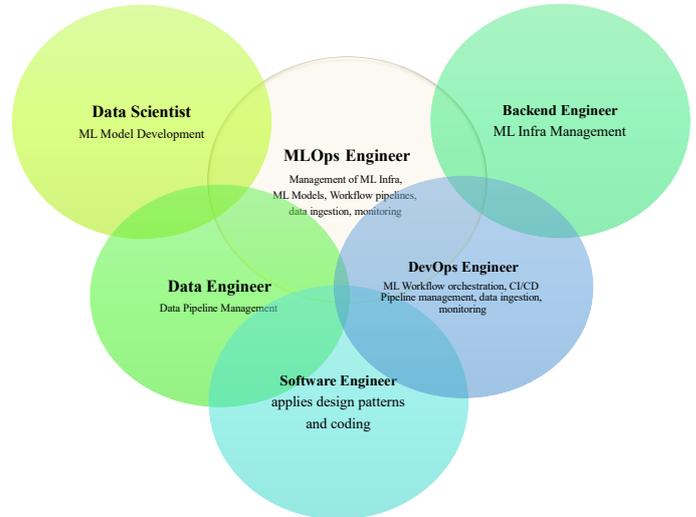

Figure 5 –MLOps Roles and Responsibilities Intersections [16]

The key advantages of MLOps include Efficiency, Scalability and Risk reduction.

1) Efficiency: MLOps facilitates faster model development by data teams, resulting in higher-quality machine learning models and expedited deployment and production processes.

2) Scalability: It allows for the extensive scalability and management of numerous models. MLOps supports continuous integration, delivery, and deployment, ensuring smooth operation and oversight.

3) Risk reduction: MLOps facilitates increased transparency and rapid response to regulatory examinations and drift evaluations of machine learning models. This approach ensures better conformity with industry's policies and standards. The image shows the steps of the machine learning process in detail [14].

Figure 6 – Machine Learning Process Step by Step [14]

## IV. MLOps Tools

The MLOps landscape featured a wide variety of tools and platforms aimed to help organizations and individuals efficiently handle parts or the entire machine learning lifecycle. The field is rapidly evolving, offering practitioners a multitude of options to effectively implement machine learning. In this part, we introduce the most important MLOps tools and platform utilized by leading technology companies in the world and mention some of their exclusive MLOps infrastructure.

A. End to End MLOps Platforms

End-to-end MLOps platforms offer a cohesive environment as a unified ecosystem that optimizes the entirety of the machine learning workflow, from data preparation and model development to deployment and monitoring [17].

Table 1 - End to End MLOps Platforms

| Name | Type | Release Date |
|---|---|---|
| Google Cloud Platform | Public | 2008 |
| Microsoft Azure | Public | 2010 |
| H2O.ai | Open Source | 2012 |
| Iguazio | Private | 2014 |
| AzureML | Open Source | 2015 |
| Databricks | Private | 2015 |
| Valohai | Public | 2016 |
| Amazon SageMaker | Public | 2017 |
| MLflow | Open Source | 2018 |
| Kubeflow | Open Source | 2018 |
| AliBaba Cloud ML | Public | 2018 |
| DataRobot | Private | 2019 |
| MetaFlow | Open Source | 2019 |
| Cloudera | Public | 2020 |
| Vertex AI | Public | 2021 |

B. Experiment tracking, metadata storage, and management

Experiment tracking and model metadata management tools enable the user to track experiment parameters, and visualizations, ensuring the ability to reproduce them and facilitating collaboration opportunities.

Table 2 - MLOps Experiment tracking, model storage, and management.

| Name | Type | Release Date |
|---|---|---|
| Neptune.ai | Private | 2017 |
| Comet ML | Private | 2017 |
| TensorBoard | Open Source | 2017 |
| Weights and Biases | Private | 2018 |
| CML | Open Source | 2018 |
| MLflow | Open Source | 2018 |
| ModelDB | Open Source | 2020 |
| AimStack | Open Source | 2021 |

C. Dataset labeling and annotation

Dataset labeling and annotation tools form a critical component of ML systems, enabling you to prepare high-quality training data for their models. These tools offer an efficient process for data annotation, guaranteeing accurate and consistent labeling that fuels model training and evaluation.

Table 3 - Dataset labeling and annotation.

| Name | Type | Release Date |
|---|---|---|
| Scale AI | Non-free | 2016 |
| Labelbox | Private | 2017 |
| Amazon Ground Truth | Public | 2018 |
| Kili | Non-free | 2018 |
| Superb AI | Non-free | 2018 |
| Snorkel Flow | Private | 2019 |
| SuperAnnotate | Non-free | 2020 |
| Encord Annotate | Non-free | 2020 |

D. Data storage and versioning

Data storage and versioning tools enables to maintain data integrity, collaboration, facilitate the reproducibility of experiments and ensure accurate ML model development and deployment. Data versioning enable to trace and compare different iterations of datasets.

Table 4 - Data storage, preprocessing and versioning.

| Name | Type | Release Date |
|---|---|---|
| iMerit | Private | 2012 |
| Pachyderm | Private | 2014 |
| Labelbox | Private | 2017 |
| Prodigy | Private | 2017 |
| Comet | Private | 2017 |
| Data Version Control | Open Source | 2017 |
| Qri | Open Source | 2018 |
| Weights and Biases | Private | 2018 |
| Delta Lake | Open Source | 2019 |
| Doccano | Open Source | 2019 |
| Snorkel | Private | 2020 |
| Supervisely | Private | 2020 |
| Segments.ai | Private | 2020 |
| Dolt | Open Source | 2020 |
| LakeFS | Open Source | 2020 |

E. Feature stores

Feature stores provide a unified framework for the storage, management, and provision of ML features. They facilitate the discovery and sharing of feature values, essential for both training and deploying machine learning models.

Table 5 – Feature Stores and Feature Engineering

| Name | Type | Year |
|---|---|---|
| Iguazio Data Platform | Private | 2014 |
| TsFresh | Private | 2016 |
| Hopsworks Feature | Open Source | 2016 |
| Featuretools | Private | 2017 |
| dotData | Private | 2018 |
| AutoFet | Open Source | 2019 |
| Feast | Open Source | 2019 |
| Tecton | Non-free | 2019 |
| Featureform | Open Source | 2019 |
| Rasgo | Private | 2020 |
| HopsWork | Private | 2021 |
| Databricks Feature | Private | 2021 |
| Vertex AI Feature Store | Public | 2021 |

F. Hyperparameter optimization

As of 2023, the landscape of hyperparameter optimization tooling has remained largely unchanged, with the familiar and established tools continuing to dominate the field.

Table 6 - Hyperparameter optimization

| Name | Type | Release Date |
|---|---|---|
| Hyperopt | Open Source | 2013 |
| SigOpt | Public | 2014 |
| Google Vizier | Public | 2017 |
| Scikit-Optimize | Open Source | 2017 |
| Optuna | Open Source | 2018 |
| Talos | Open Source | 2018 |
| Optuna | Open Source | 2019 |

G. Workflow orchestration and pipelining tools

Workflow orchestration and pipelining tools are essential components for streamlining and automating complex ML workflows.

Table 7 - Workflow orchestration and pipelining tools

| Name | Type |
|---|---|
| ZenML | Open Source |
| Kedro Pipelines | Open Source |
| Flyte | Open Source |
| Prefect | Open Source |
| Mage AI | Open Source |

### H. Model deployment and serving

Model deployment and serving tools enable you to deploy trained models into production environments and serve predictions to end-users or downstream systems.

Table 8 - Model deployment and serving.

| Name | Type | Release Date |
|---|---|---|
| Algorithmia | Private | 2014 |
| TensorFlow Serving | Open Source | 2016 |
| KubeFlow | Open Source | 2018 |
| OpenVino | Open Source | 2018 |
| Triton Inference Server | Open Source | 2018 |
| BentoML | Open Source | 2019 |
| OctoML | Open Source | 2019 |
| Seldon Core | Private | 2020 |
| Torch Serve | Open Source | 2020 |
| KFServing | Open Source | 2020 |
| Syndicai | Private | 2020 |
| BodyWork | Open Source | 2021 |
| Cortex | Private | 2021 |
| Sagify | Open Source | 2021 |

### I. Model observability

Model observability tools can allow you to gain insights into the behavior, performance, and health of your deployed ML models.

Table 9 - Model observability

| Name | Type | Release Date |
|---|---|---|
| Unravel Data | Private | 2013 |
| Fiddler AI | Private | 2018 |
| Losswise | Private | 2018 |
| Superwise | Private | 2019 |
| MLrun | Open Source | 2019 |
| WhyLabs | Open Source | 2020 |
| Arize AI | Private | 2020 |
| Evidently AI | Open Source | 2020 |
| Aporia | Open Source | 2021 |
| Deep Checks | Private | 2021 |

### J. The Massive Industrial Companies Cases

In the recent years, some of the most leading technology companies have launched their own dedicated MLOps platforms while they have faced two fundamental challenges: First, the time needed on design, build, and deploy the ML model in the operational environment. The main goal is to reduce the time from several months to several weeks. Also, the stability of models in terms of predictions and reproduction of these models in various and complex conditions are the most important goals of leading companies such as Netflix [18][19], Uber [20][21][22], Databricks [23][24], Google [25], Airbnb [26], and other companies such as Walgreens Boots Alliance [27], DoorDash [28] and Spotify [29][30].

## V. EXPERTS EXPERIENCES

In this study, over 110 experts participated in an online questionnaire and answered 14 questions. After weeks of collecting feedback from experts, we started to analyze the answers. The design of the questions has been done in consultation with several experts in the field. Targeted professionals are MLOps Engineers, ML Engineers, Data Scientists, Software Engineers, Data Engineers, DevOps Engineers, Backend Engineers, and AI Engineers.

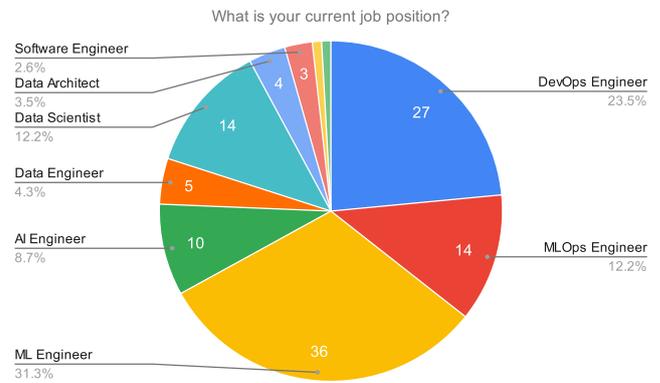

Figure 3 – Experts Current Job Positions

Figure 3 shows the current job position of our respondents. The results show that, 36 (31.3%) are Machine Learning Engineer, 27 (23.5%) are DevOps Engineer, 14 (12.2%) are MLOps Engineer, 14 (12.2%) are Data Scientist,10 (8.7%) is Artificial Intelligence Engineer, 5 (4.3%) are Data Engineer, and 3 (2.6%) are Software Engineer.

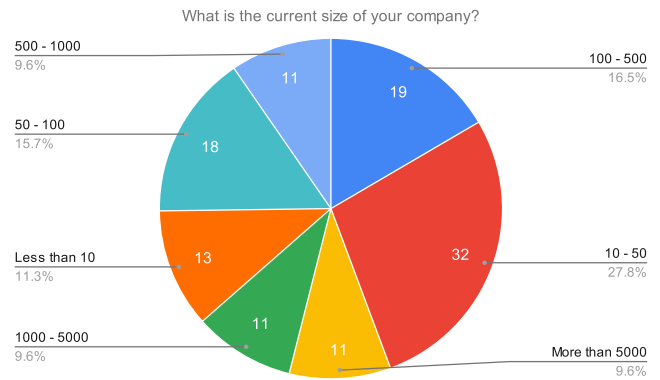

Figure 4 –Size of Companies

Figure 4 shows the size of our respondent's company. The results show that, approximately 32 individuals (27.8%), are employed in organizations ranging in size from 10 to 50 employees, 19 individuals (16.5%), are employed in organizations ranging in size from 100 to 500 employees, 18 individuals (15.7%), are employed in organizations ranging in size from 50 to 100 employees, 13 individuals (11.3%), are employed in organizations ranging in size less than 10 employees, 11 individuals (9.6%), are employed in organizations ranging in size from 1000 to 50000 employees, 11 individuals (9.6%), are employed in organizations ranging in size more than 5000 employees, and 11 individuals (9.6%), are employed in organizations ranging in size more from 500 to 1000 employees,

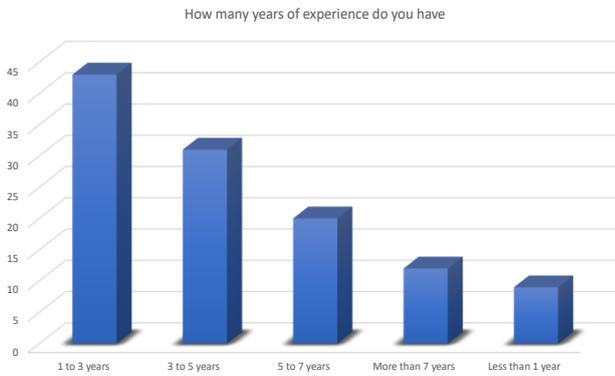

Figure 5 – Experts Experiences per Year

Figure 5 displays the quantity of individuals' years of experience. Based on the data analysis, our report indicates that 43 individuals (37.4%) possess a work experience ranging from 1 to 3 years, 31 individuals (27%) possess a work experience ranging from 3 to 5 years, 20 individuals (17.4%) possess a work experience ranging from 5 to 7 years, 11 individuals (9.6%) possess a work experience more than 7 years, 9 individuals (7.8%) possess a work experience less than a year and 1 individual (0.9%) possess a work experience more than 15 years.

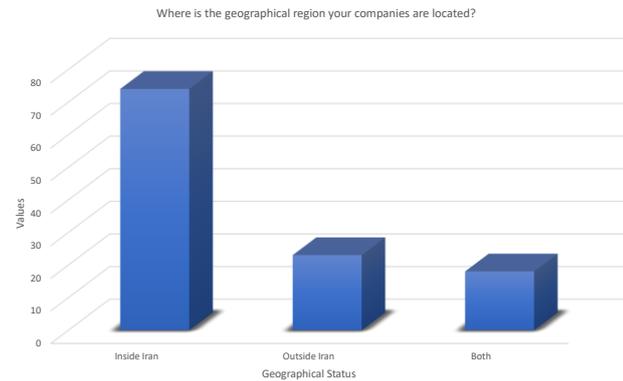

Figure 6 – Companies Geographical Location

Figure 6 shows the geographical distribution of companies to which individuals have affiliations. The results indicate that 74 individuals (64.3%) are employed by companies located within Iran, 23 individuals (20%) are affiliated with companies outside of Iran, and 18 people (15.7%) are engaged in collaborative work with both domestic and international companies.

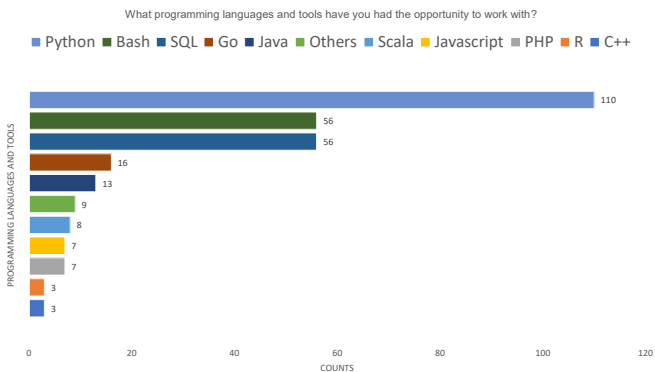

Figure 7 - Programming Languages

Figure 7 displays the programming languages in which individuals have the highest coding and software development experience. The results indicate that 110 programmers (95.7%) primarily utilize the Python programming language for their coding and software development, total of 56 individuals (48.7%) employ SQL for tasks related to database topics, 56 individuals (48.7%) seek assistance from the Bash command language within the Unix and Linux environments, 16 individuals (13.9%) find greater assistance from the statically typed, compiled high-level Go programming language, 13 individuals (11.3%) uses the Java programming language for software development efforts, 8 individuals (7%) use strong statically typed high-level general-purpose Scala programming language for coding, 7 individuals (6.1%) uses the JavaScript for web development tasks, 3 individuals utilize the R language for statistical data analysis, while another three people employ the C++ language for the development of system programs, and 9 individuals use other programming languages and tools for their specific software development tasks.

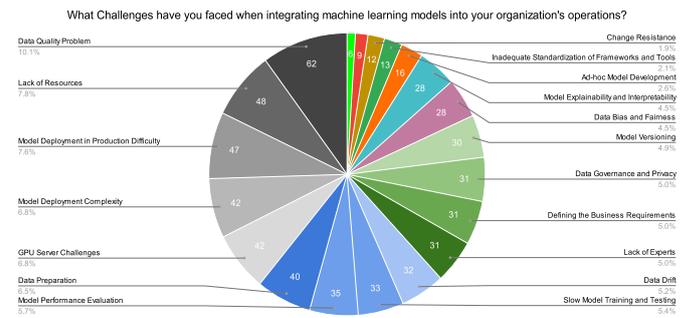

Figure 8 – MLOps Different Challenges

Figure 8 refers to the challenges and difficulties faced by developers in implementing the MLOps paradigm at Enterprise. The table 10 lists the challenges in descending order of importance, with the most significant challenges at the top and the less critical ones toward the bottom.

Table 10 - MLOps Challenges in order of Importance at Enterprise

| Challenges | Votes |
| --- | --- |
| Data Quality Problem | 62 |
| Lack of Resources | 48 |
| Model Deployment in Production Difficulty | 47 |
| GPU Server Challenges | 42 |
| Model Deployment Complexity | 42 |
| Data Preparation | 40 |
| Model Performance Evaluation | 35 |
| Slow Model Training and Testing | 33 |
| Data Drift | 32 |
| Data Governance and Privacy | 31 |
| Defining the Business Requirements | 31 |
| Lack of Experts | 31 |
| Model Versioning | 30 |
| Model Explainability and Interpretability | 28 |
| Data Bias and Fairness | 28 |
| Ad-hoc Model Development | 16 |
| Inadequate Standardization of Frameworks and Tools | 13 |
| Change Resistance | 12 |
| Others | 9 |
| Data Silos and Fragmentation | 6 |

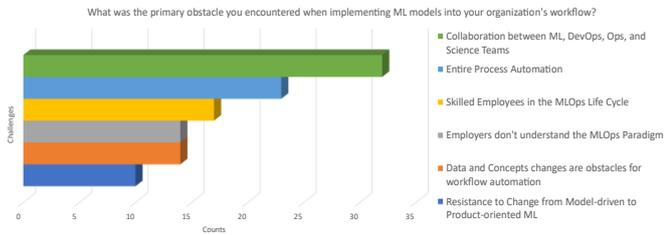

Figure 9 - MLOps Primarily Obstacles

Figure 9 shows the primary and pivotal challenge of MLOps within the organization as determined by experts' viewpoints while in the table 11 Primary Challenge are listed by expert's viewpoints.

Table 11 - Primary Challenge of MLOps at Enterprise

| Challenges | Votes |
| --- | --- |
| Collaboration between ML, DevOps, Ops, and Science Teams | 32 |
| Entire Process Automation | 23 |
| Skilled Employees in the MLOps Life Cycle | 17 |
| Data and Concepts changes are obstacles for workflow automation | 14 |
| Employers don't understand the MLOps Paradigm | 14 |
| Resistance to Change from Model-driven to Product-oriented ML | 10 |

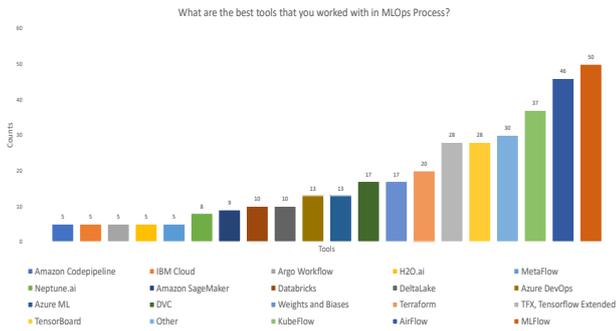

Figure 10 - Most frequently utilized MLOps Tools at Enterprise

Figure 10 refers to the Most frequently utilized MLOps Tools at Enterprise, while in the table 12, tools are listed by popularity order.

Table 12 - Most frequently utilized MLOps Tools at Enterprise

| Tools | Counts |
| --- | --- |
| MLFlow | 50 |
| AirFlow | 46 |
| KubeFlow | 37 |
| Apache Kafka | 31 |
| TFX, Tensorflow Extended | 28 |
| TensorBoard | 28 |
| Ansbile | 22 |
| Terraform | 20 |
| DVC | 17 |
| Weights and Biases | 17 |
| Azure DevOps | 13 |
| Azure ML | 13 |
| Databricks | 10 |
| DeltaLake | 10 |
| Microsoft PowerShell | 10 |
| Amazon SageMaker | 9 |
| Neptune.ai | 8 |
| Amazon Codepipeline | 5 |
| IBM Cloud | 5 |
| Argo Workflow | 5 |
| H2O.ai | 5 |
| MetaFlow | 5 |

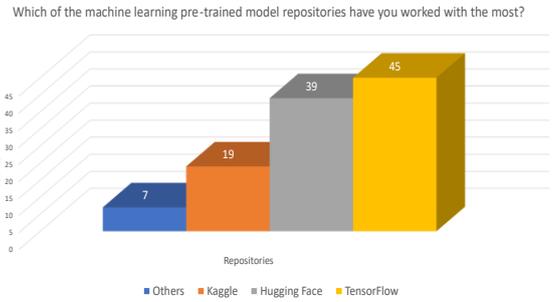

Figure 11 - Pre-trained Machine Learning Repositories

The findings indicate that among the 110 respondents, 45 individuals primarily relied on the TensorFlow repository for pre-trained machine learning models, 39 people engaged with the Hugging Face repository, 19 respondents predominantly utilized the Kaggle repository, and 7 individuals used other repositories.

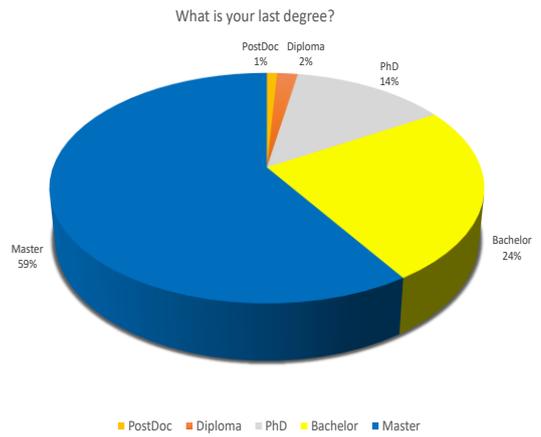

Figure 12 - Experts Educational Levels

The results show that out of 110 respondents, 65 (59%) have a master's degree, 27 (24%) have a bachelor's degree, 15 (14%) have a PhD, 2 (2%) have a diploma and 1 (1%) is a Postdoctoral researcher.

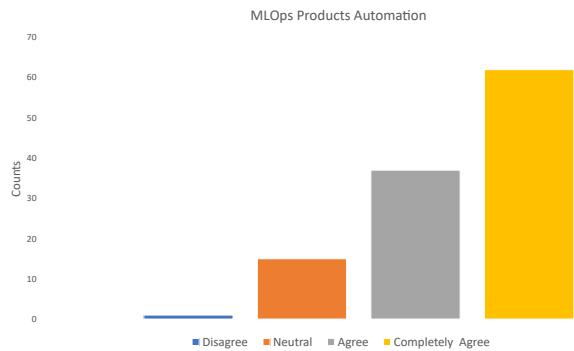

Figure 13 - MLOps Automation

Figure 14 shows the opinions of experts regarding the following phrase: "MLOps is closely associated with the collaboration between ML engineers and developers to automate the development of ML products". The results show that out of 110 respondents, 60 (53.9%) agree with the statement, 35 (32.2%) completely agree, 14 (13%) are neutral, and 1 (0.9%) disagree with the statement.

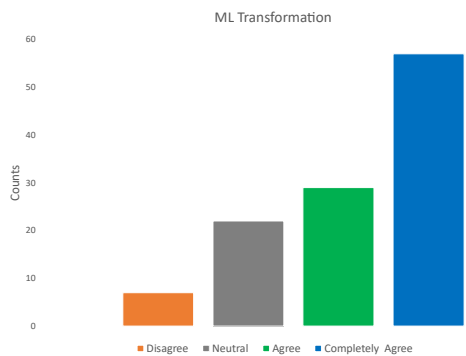

Figure 14 – Machine Learning Transformation

Figure 14 shows the experts opinions regarding the following phrase: "MLOps is currently undergoing a significant transformation in ML Engineering and revolutionizing the way in which these processes are conducted". The results indicates out of 110 respondents, 55 (49.6%) agree with the statement, 22 (19.1%) completely agree, 27 (25.2%) are neutral, and 6 (6.1%) disagree with the statement.

## VI. CONCLUSION

This study has presented a brief overview of the evolving landscape of MLOps and its significance in addressing the challenges faced by enterprises in managing machine learning models effectively. We presented an extensive list of MLOps tools and platforms utilized by leading technology companies, offering readers a valuable resource for navigating the MLOps landscape. Furthermore, the research has incorporated insights from over 110 Iranian Software experts, providing a glimpse into their experiences, challenges, and preferences. The findings reveal that data quality problems, a lack of resources, and difficulties in model deployment are among the primary challenges faced by practitioners. Collaboration between ML, DevOps, Ops, and Science teams is seen as a pivotal challenge in implementing MLOps effectively.

## ACKNOWLEDGMENT

The authors, express their endless gratitude to all Iranian machine learning activists and software developers for sharing their valuable experiences towards conducting this research.